
\magnification 1200
\tenpoint
\superrefsfalse
\baselineskip = 0.333 truein \relax
\def\frac#1#2{{#1\over#2}}

\hskip 1cm {\bf Gauge invariant three-boson vertices in the Standard Model}
\vskip 0.2cm
\hskip 3.5cm  {\bf and the static properties of the W.}
\vskip 0.8cm
\hskip 2.7cm {\bf Joannis Papavassiliou and Kostas Philippides}
\vskip 0.2cm
\hskip 3cm Department of Physics, New York University,
\vskip 0.1cm
\hskip 3cm 4 Washington Place, New York, N.Y. 10003

\vskip 0.5cm

\hskip 5cm {\bf ABSTRACT}
\vskip 0.2cm

We use the S-matrix pinch technique to derive to one-loop order
gauge-independent
 $\gamma W^{+}W^{-}$ and $ZW^{+}W^{-}$
 vertices in the context of the Standard Model,
with all three incoming momenta off-shell. We show that
 the $\gamma W^{+}W^{-}$ vertex so constructed is related to the
gauge-independent $W$ self-enery, derived by Degrassi and Sirlin, by
a very simple QED-like Ward identity. The same results
 are obtained by
the pinch technique applied directly to the process $e^{+}e^{-}
\rightarrow W^{+}W^{-}$. Explicit calculations
 give rise to
expressions for static properties of the W gauge bosons
like magnetic dipole and electric quadrupole moments,
which satisfy the crucial properties of
infrared finiteness
as well as gauge-independence.

\vskip 5cm

\section {Introduction}

The widespread belief of theoreticians that the Standard Model of
electro-weak interactions serves as an effective
low energy description of an
underlying more fundamental theory
has been persistently denied confirmation for several years, mainly
 because of the Model's
impressive agreement with a large body of precise experimental results
\reference{SUSY93}
A.~Sirlin, Talk at the SUSY 93 International Workshop,
Northeastern University.
\endreference
{}.
This fact, combined with the elusiveness of the top and Higgs particles,
has motivated the design of a new generation of machines, which will
further probe the underpinnings of the Standard Model in the near future.
A new and largely unexplored frontier, on which this ongoing search for new
Physics will soon focus, is the study of the structure of the three-boson
couplings. Since
these couplings lie in the heart of the non-Abelian nature
of the theory, a systematic confrontation
of the Standard Model predictions in this domain with experiment, might
shed some light on the underlying structure of the theory, if any.

Motivated by scattering experiments of the form
 $e^{+} e^{-} \rightarrow W^{+} W^{-}$, the standard parametrization
for the most
general $W^{+}W^{-}V$ vertex with the W's on shell and V off-shell,
where V stands for $\gamma$ or Z, is as follows
\reference{Goun1}
K.~J.~F.~Gaemers and G.~J.~Gounaris
\journal Z. Phys.; C 1,259 (1979)
\endreference
\reference{Hagi}
K.~Hagiwara et {\sl al},
\journal Nucl. Phys.; B 282,253 (1987)
\endreference
\reference{Z1}
U.~Baur and D.~Zeppenfeld,
\journal Nucl. Phys.; B 308,127 (1988)
\endreference
\reference{Z2}
U.~Baur and D.~Zeppenfeld,
\journal Nucl. Phys.; B 325,253 (1989)
\endreference
\reference{Arg1}
E.~N.~Argyres et {\sl al},
\journal Phys. Lett.; B 272,431 (1991)
\endreference
\reference{Lahanas}
E.~N.~Argyres et {\sl al},
\journal Nucl. Phys.; B 391,23 (1993)
\endreference
:
$$\eqalign{
\Gamma_{\mu\alpha\beta}^{V}=& -ig_{V} \Biggl\lbrack
 f[2g_{\alpha\beta}\Delta_{\mu}+ 4(g_{\alpha\mu}Q_{\beta}-
g_{\beta\mu}Q_{\alpha})]\cr
& + 2\Delta\kappa_{V}(g_{\alpha\mu}Q_{\beta}-g_{\beta\mu}Q_{\alpha})\cr
& + 4\frac{\Delta Q_{V}}{M_{W}^{2}}(\Delta_{\mu}Q_{\alpha}Q_{\beta}-
\frac{1}{2}Q^{2}g_{\alpha\beta}\Delta_{\mu})\Biggr\rbrack + ...\cr}
\EQN Parametrization$$
with $g_{\gamma}= gs$ and $g_{Z}= gc$, where $g$ is the $SU(2)$
gauge coupling
and $s\equiv sin\theta_{w}$ and $c\equiv cos\theta_{w}$,
 and the dots
denote omission of C,P or T violating terms.
 The four-momenta $Q$ and $\Delta$, first introduced in
\reference{Bardeen}
W.~A.~Bardeen, R.~Gastmans, and B.~Lautrup,
\journal Nucl. Phys.; B 46,319 (1972)
\endreference
, are related to the incoming momenta $q$, $p_{1}$ and $p_{2}$ by
$q=2Q$, $p_{1}=-\Delta -Q$ and $p_{2}=\Delta - Q$, as shown in Fig.1 .
The quantities $\Delta\kappa_{V}$ and $\Delta Q_{V}$
are defined as
$$
\Delta\kappa_{V}= \kappa_{V} + \lambda_{V} - 1
\EQN deltakappa$$
and
$$
 \Delta Q_{V}= -2\lambda_{V}
\EQN deltaQ$$

where $\kappa_{V}$ and $\lambda_{V}$ are form factors compatible with C, P,
and T invariance.
 In particular, $\kappa_{\gamma}$ and $\lambda_{\gamma}$ are
related to the magnetic dipole moment $\mu_{W}$ and the electric
quadrupole moment $Q_{W}$, by the following expressions:
$$
\mu_{W}= \frac{e}{2M_{W}}(1+\kappa_{\gamma}+\lambda_{\gamma})
\EQN dipole$$
and
$$
Q_{W}=- \frac{e}{M^{2}_{W}}(\kappa_{\gamma} - \lambda_{\gamma})
\EQN quad$$
In the context of the Standard Model,
 the tree level values of the parameters defined above
 are $f=1$, $\Delta\kappa_{V}=\Delta Q_{V}=0$, $\kappa_{\gamma}=1$ and
$\lambda_{\gamma}=0$.
 Given the expected experimental precision, and assuming that such quantities
can actually be extracted from
suitable scattering experiments,
 calculating the one-loop
corrections to these parameters is the next necessary step.
One then must cast the resulting expressions in the following form:
$$
\Gamma_{\mu\alpha\beta}^{V}= -ig_{V}[
 a_{1}^{V}g_{\alpha\beta}\Delta_{\mu}+ a_{2}^{V}(g_{\alpha\mu}Q_{\beta}-
g_{\beta\mu}Q_{\alpha})
 + a_{3}^{V}\Delta_{\mu}Q_{\alpha}Q_{\beta}]
\EQN 1loopParametrization$$
where $a_{1}^{V}$, $a_{2}^{V}$, and $a_{3}^{V}$ are
 in general complicated functions of the
momentum transfer $Q^2$ and the masses of the particles appearing in the
loops.
It then follows that $\Delta\kappa_{V}$ and $\Delta Q_{V}$ are given by the
following expressions:
$$
\Delta\kappa_{V}=\frac{1}{2}(a_{2}^{V}-2a_{1}^{V}-Q^{2}a_{3}^{V})
\EQN 1loopdeltakappa$$
and
$$
\Delta Q_{V}= \frac{M^{2}_{W}}{4}a_{3}^{V}
\EQN 1loopdeltaQ$$

Calculating the one-loop expressions for $\Delta\kappa_{V}$ and
 $\Delta Q_{V}$ is a non-trivial task, not only from the technical point of
view, but from the conceptual point of view as well.
So, if one proceeded to
calculate
just the Feynman diagrams contributing to the $\gamma W^{+}W^{-}$
vertex, for example, and then extract from them the contributions to
$\Delta\kappa_{\gamma}$ and $\Delta Q_{\gamma}$,
 after a considerable amount of labor one would arive at
expressions that are
plagued with several pathologies, gauge-dependence being one of them.
Indeed, even if the two W are considered to be on shell, since the incoming
photon is not, there is no {\sl a priori}
 reason why a gauge-independent answer
should emerge. In the context of the renormalizable gauges
(usually referred to as $R_{\xi}$ gauges) the final answer depends on the
choice of the gauge fixing parameter $\xi$, which enters into the one-loop
calculations through the gauge-boson propagators
( W,Z,$\gamma$, and unphysical Higgs particles).
In addition, as it was realized
 by the authors of \cite{Lahanas}, who,
unaware of the fact that there is no gauge cancellation, performed
the calculation
in the Feynman-t'Hooft gauge ($\xi=1$),
 the answer also turns out to be
{\sl infrared divergent}.
Put in the standard language, the functions $a_{i}(Q)$ emerging out of
this procedure are in general
 gauge-dependent and infrared divergent, and these
problems persist {\sl even} when one combines them
to form  $\Delta\kappa_{\gamma}$ and $\Delta Q_{\gamma}$, according to
\Eq{1loopdeltakappa} and \Eq{1loopdeltaQ} (the resulting expressions are
however ultraviolet finite).
 Clearly, regardless of the measurability of
quantities like $\Delta\kappa_{\gamma}$ and $\Delta Q_{\gamma}$, from
the theoretical point of view one should at least be able to satisfy
such crucial requirements as gauge-independence and infrared finiteness,
when calculating the model's prediction for them.

This unsatisfactory state of affairs,
which occurs each time one tries to
isolate a particular piece of an S-matrix without enough care, can be avoided
if one adopts the pinch technique (P.T.).
The P.T. was invented by Cornwall over a decade ago
\reference{Kursunoglu}
J.~M.~Cornwall,
in Deeper Pathways in High Energy Physics, edited by B.~Kursunoglu,
A.~Perlmutter, and L.~Scott (Plenum, NewYork, 1977), p.683
\endreference
and has since been applied to
a variety of physical problems. The main idea of this method
is to resum via a well-defined algorithm the Feynman diagrams
contributing to a gauge-invariant process
(like an S-matrix element), in such a way as to form new
gauge-independent proper vertices, and new propagators with
gauge-independent self-energies and only a trivial gauge dependence
- that of their tree level parts.
 In the context of QCD a gauge invariant
gluon self-energy was derived, and its Schwinger-Dyson
 equation constructed and
solved for $T=0$
\reference{Bible}
J.~M.~Cornwall,
\journal Phys. Rev.; D 26,1453 (1982)
\endreference
, as well as finite $T$
\reference{Hou}
J.~M.~Cornwall, W.~S.~Hou, and J.~E.~King,
\journal Phys. Lett.; B 153,173 (1988)
\endreference
 . The plasmon decay rate was also calculated at finite $T$ using the same
method
\reference{Nadkarni}
S.~Nadkarni,
\journal Phys. Rev. Lett; 61,396 (1988)
\endreference
. Later the QCD gauge invariant three-gluon vertex was
 calculated at one
loop level and was shown to satisfy a very simple Ward
 identity
\reference{Mike and I}
J.~M.~Cornwall and J.~Papavassiliou,
\journal Phys. Rev.; D 40,3474 (1989)
\endreference
. The subleading corrections to the self-energy were calculated
by Lavelle
\reference{Lavelle}
M.~Lavelle,
\journal Phys. Rev.; D 44,26 (1991)
\endreference
{}.
Finaly, the gauge-invariant four-gluon QCD
 vertex was constructed and its
Ward identity derived
\reference{teleftaio}
J.~Papavassiliou,
\journal Phys. Rev.; D 47,4728 (1993)
\endreference
. The P.T. was first extended to the case of non-Abelian gauge theories
with spontaneously broken gauge symmetry (with elementary Higgs) in
the context of a toy field theory based on $SU(2)$,
and a gauge independent electromagnetic form factor for the
Standard Model neutrino was constructed
\reference{Klassiko}
J.~Papavassiliou,
\journal Phys. Rev.; D 41,3179 (1990)
\endreference
. The complicated task of applying the P.T. in the
electro-weak sector of the
 Standard Model was recently accomplished
 by Degrassi and
 Sirlin
\reference{Degrassi and Sirlin}
G.~Degrassi and A.~Sirlin,
\journal Phys. Rev.; D 46,3104 (1992)
\endreference
{}.
These last authors, in addition to deriving
explicit expressions for the one loop
gauge-invariant $WW$ and $ZZ$ self-energies, introduced an alternative
description of P.T. in terms of
equal time commutators of currents.

In this paper we use the
S-matrix P.T. to construct in the context of the
Standard Model
\reference{Rujula}
In this paper we do not touch upon issues
related to the gauge invariance of
effective trilinear couplings stemming from theories beyond
the Standard Model.
\endreference
, to one-loop order in perturbation theory, a
 {\sl gauge-invariant} effective
$\gamma WW$ vertex with {\sl all three} incoming momenta being
{\sl off-shell}. The outline of the construction of such a vertex was
already given in \cite{Degrassi and Sirlin}, but no explicit results were
reported. It turns out that the vertex
${\hat{\Gamma}}_{\mu\alpha\beta}(q,p_{1},p_{2})$
 so constructed satisfies a very simple
QED-like Ward identity, which relates it to the {\sl gauge independent}
WW self-energy ${\hat{\Pi}}_{\alpha\beta}$
derived in \cite{Degrassi and Sirlin}, namely
$$
 q^{\mu}{\hat{\Gamma}}_{\mu\alpha\beta}
 = {\hat{\Pi}}_{\alpha\beta}(p_{1})
                                  -{\hat{\Pi}}_{\alpha\beta}(p_{2})
\EQN MajorWI$$
where $q$ is the four-momentum of the incoming photon, and $p_1$ and
$p_2$ of the incoming $W^{+}$ and $W^{-}$ respectively.
Having constructed a gauge-independent vertex for the general case of
off-shell momenta we can recover as a special limit the case of interest
($W^{+}$ and $W^{-}$ on shell)
by setting $p^{2}_{1} \rightarrow M^{2}_{W}$
and $p^{2}_{2} \rightarrow M^{2}_{W}$ and contracting the result with the
polarization vectors $\epsilon^{\alpha}(p_{1})$ and
$\epsilon^{\beta}(p_{2})$. Finally, projecting out the kinematically
relevant pieces according to \Eq{1loopParametrization} gives rise to
new individually
 {\sl gauge-independent}
and {\sl infrared finite} functions ${\hat{a}}_{i}(Q^2)$.
So, the additional (ultraviolet finite) pinch
contributions not only contain the right terms to cancel all gauge
dependences, but they also contribute infrared divergent terms, which
{\sl exactly} cancel against the infrared divergences contained in the
standard vertex graphs. Combining now the new functions
 ${\hat{a}}_{i}(Q^2)$ according to \Eq{1loopdeltakappa} and
\Eq{1loopdeltaQ}
 we find expressions for
 $\Delta\kappa_{\gamma}$ and $\Delta Q_{\gamma}$ that are

1) Gauge fixing parameter ($\xi$) independent

2) Ultraviolet {\sl and} infrared finite.

3) Well behaved for large momentum transfer ($Q^{2}$)

The paper is organized as follows. In section 2 we review the
S-matrix P.T. and discuss some of the more important results for our
purposes. In addition, we briefly present Degrassi's and Sirlin's
alternative formulation
of the P.T. In section 3 the S-matrix P.T. is used to construct the
gauge-independent $\gamma W^{+}W^{-}$ vertex with {all} incoming momenta
off-shell, and explicit results are reported. Furthermore, we present a
short-cut for deriving the gauge-independent vertex with the two W on-shell
{\sl directly} from a process like $e^{+} e^{-} \rightarrow W^{+} W^{-}$.
In section 4 we prove the Ward identity that the gauge independent vertex
satisfies. In section 5 we calculate the functions
${\hat{a}}(Q^{2})$ and
evaluate numerically the pinch contributions to
$\Delta\kappa_{\gamma}$ and $\Delta Q_{\gamma}$ .
Finally, we summarise our results in section 6.

\section {The Pinch Technique.}
In this section we briefly review the S-matrix pinch technique (P.T.).
 In particular we outline the method of derivation of
 the gauge-independent proper self-energy of a gauge boson
 and comment on the technical
differences that arise
 when P.T. is applied to a theory with symmetry breaking,
like the electro-weak theory, as opposed to a theory like QCD.
 In addition,
we present the main idea of Degrassi's
 and Sirlin's formulation of the P.T.,
and establish some of the notation
 we will use in the sequel.

The S-matrix pinch technique is an algorithm that allows the construction
of modified gauge independent n-point functions, through the
order by order resummation of
Feynman graphs contributing to a certain physical
and therefore ostensibly gauge independent process
(an S-matrix in our case).
The simplest example that demonstrates how the P.T. works is the gauge boson
two point function (propagator).
Consider the $S$-matrix
element $T$ for the elastic scattering of two fermions of masses
$M_{1}$ and $M_{2}$. To any order in perturbation theory $T$ is independent
of the gauge fixing parameter $\xi$, defined by the free
gluon propagator
$$
\Delta_{\mu\nu}(q) = \frac{-g_{\mu\nu}+(1-\xi)q_{\mu}q_{\nu}/q^2}
                          {q^2}
\EQN GaugeProp$$
On the other hand, as an explicit calculation shows,
the conventionally defined proper self-energy (collectively
depicted in graph 2a)
depends on $\xi$. At the one loop level this dependence is canceled by
contributions from other graphs, like 2b and 2c, which do not seem to be
of propagator type at first glance.
That this must be so is evident from the form of $T$:
$$
T(s,t,M_{1},M_{2})= T_{1}(t) + T_{2}(t,M_{1},M_{2})+T_{3}(s,t,M_{1},M_{2})
\EQN S-matrix$$
where the function $T_{1}(t)$ depends only on the Mandelstam variable
$t=-({\hat{p}}_{1}-p_{1})^{2}=-q^2$,
 and not on $s=(p_{1}+p_{2})^{2}$ or on the
external masses.
  The propagator-like parts of graphs like 1e and 1f,
which enforce the gauge independence of $T_{1}(t)$,
 are called  "pinch parts".
The pinch parts emerge every time a gluon propagator or an elementary
three-gluon vertex contribute a longitudinal $k_{\mu}$ to the original
graph's numerator. The action of such a term is
to trigger an elementary
Ward identity of the form
$$\eqalign{
k^{\mu}\gamma_{\mu} \equiv & \slashchar{k} = (\slashchar{p}+
\slashchar{k}-m)-(\slashchar{p}-m)\cr
=& S^{-1}(p+k)-S^{-1}(p)\cr}
\EQN BasicPinch$$
once it gets contracted with a $\gamma$ matrix.
The first term on the right-hand side of \Eq{BasicPinch} will remove the
internal fermion propagator - that is a "pinch" - whereas $S^{-1}(p)$
vanishes
on shell. This last property characterizes the S-matrix P.T. we will use
throughout this paper.
Returning to the decomposition of \Eq{S-matrix}, the function $T_{1}(t)$
is gauge invariant and unique and represents the contribution
of the new propagator.
We can construct the new propagator, or equivalently $T_{1}(t)$, directly
from the Feynman rules.
 In doing so it is evident that any value for the gauge
parameter $\xi$ may be chosen, since $T_{1}$, $T_{2}$, and $T_{3}$ are
all independent of $\xi$. The simplest of all covariant gauges is
certainly the Feynman-t'Hooft gauge ($\xi = 1$),
 which removes the
longitudinal part of the gluon propagator. Therefore, the only possibility
for pinching arises from the
four-momentum of the three-gluon vertices, and the only
 propagator-like contributions come from graph 2b.

To explicitly calculate the pinching contribution of a graph such as 2b
it is convenient to decompose the vertex in the following way, first
proposed by 't Hooft. Group theory factors aside,
$$
\Gamma_{\mu\nu\alpha}=\Gamma^{P}_{\mu\nu\alpha} + \Gamma^{F}_{\mu\nu\alpha}
\EQN tHooft$$
with
$$
\Gamma^{P}_{\mu\nu\alpha} \equiv (q+k)_{\nu}g_{\mu\alpha}
 + k_{\mu}g_{\nu\alpha}
\EQN GammaP$$
and
$$
\Gamma^{F}_{\mu\nu\alpha} \equiv 2q_{\mu}g_{\nu\alpha} -
 2q_{\nu}g_{\nu\alpha} - (2k+q)_{\alpha}g_{\mu\nu}
\EQN GammaF$$

$\Gamma^{F}_{\mu\nu\alpha}$ satisfies a Feynman-gauge Ward identity:

$$
q^{\alpha}\Gamma^{F}_{\mu\nu\alpha} = [k^2-(k+q)^2]g_{\mu\nu}
\EQN FeynWard$$
where the RHS is the difference of two inverse propagators.
As for $\Gamma^{P}_{\mu\nu\alpha}$ (P for "pinch") it gives rise to pinch
parts when contracted with $\gamma$ matrices
$$\eqalign{
g_{\mu\alpha}(\slashchar{q}+\slashchar{k})=& g_{\mu\alpha}[(\slashchar{p}
+\slashchar{q}-m)-(\slashchar{p}-\slashchar{k}-m)]\cr
=& g_{\mu\alpha}[S^{-1}(p+q)-S^{-1}(p-k)]\cr}
\EQN Pinch1$$
and
$$\eqalign{
g_{\nu\alpha}\slashchar{k}=& g_{\nu\alpha}[(\slashchar{p}-m)
-(\slashchar{p}-\slashchar{k}-m)]\cr
=& g_{\nu\alpha}[S^{-1}(p)-S^{-1}(p-k)]\cr}
\EQN Pinch2$$
Now both
$S^{-1}(p+q)$ and $S^{-1}(p)$ vanish on shell, whereas the two terms
proportional to $S^{-1}(p-k)$ pinch out the internal fermion propagator
in graph 2b.
The total pinch contribution $\Pi^{P}(q)$
 from graph 2b and its counterpart mirror
image graph with the bubble attached to the left line is given by:
$$\eqalign{
\Pi^{P}(q)=&(\frac{1}{2}N)\times 2 \times 2 \times [\frac{ig^2}{(2\pi)^4}
\int\mathinner{\frac{d^4k}{k^2(k+q)^2}}\cr
=& -\frac{2Ng^2}{16\pi^2}\ln(\frac{-q^2}{\mu^2})\cr}
\EQN TotalPinch$$
where in the second equality we give the renormalized version of the
integral
The factors in front of the integral are a group-theoretic factor
 $\frac{1}{2}N$ [ $N$ = number of colors in $SU(N)$ ]; one factor of 2 fo
the two pinching terms from \Eq{Pinch1} and \Eq{Pinch2}; another factor
of 2 from the contribution of the mirror graph.
In adding the pinch parts to the usual gluon self-energy one ambiguity
needs resolution. Because we are working with the on-shell S-matrix, any
terms $\sim q_{\mu}q_{\nu}$ in the pinch parts do not show up in
$T_{1}(t)$. We define uniquely the proper self-energy associated with the
pinch parts by demanding that it be conserved
\reference{african shield}
This conserved form is, in fact, automatic in other forms of the pinch
technique, e.g., the off-shell approach of Refs \cite{Kursunoglu} and
\cite{Bible}, or the intrinsic pinch discussed in Ref \cite{Mike and I}
\endreference
 . So we define
$\Pi^{P}_{\mu\nu}(q)$ as
$$
\Pi^{P}_{\mu\nu}(q)= P_{\mu\nu}(q)\Pi^{P}(q)
\EQN TransPinch$$
where
$$
P_{\mu\nu}(q) \equiv  -q^{2}g_{\mu\nu} + q_{\mu}q_{\nu}
\EQN ProjOper$$
Adding this to the usual Feynman-gauge proper self-energy
$$
\Pi^{(\xi=1)}_{\mu\nu}(q)= \Pi^{(\xi=1)}(q) P_{\mu\nu}(q)
\EQN FeynGaugeTens$$
with
$$
\Pi^{(\xi=1)}(q)= -\frac{5}{3}N\frac{g^2}{16\pi^2}
\ln(\frac{-q^2}{\mu^2})
\EQN FeynGaugeScal$$
we find for $\hat{\Pi}_{\mu\nu}(q)$ the gauge invariant combination:
$$
\hat{\Pi}_{\mu\nu}(q)= P_{\mu\nu}(q)\hat{\Pi}(q)
\EQN TransProp$$
with
$$
\hat{\Pi}(q)= - bg^2\ln(\frac{-q^2}{\mu^2})
\EQN RunnCoupl$$
and
$$
b= \frac{11N}{48\pi^2}
\EQN Beta$$
the coefficient of $- g^3$ in the usual one loop $\beta$ function.
Finally, the full modified propagator $\hat{\Delta}_{\mu\nu}(q)$
 at one-loop order reads
$$
\hat{\Delta}_{\mu\nu}(q)= P_{\mu\nu}(q)\hat{d}(q) -
 \xi\frac{q_{\mu}q_{\nu}}{q^4}
\EQN FullProp$$
with
$$\eqalign{
\hat{d}^{-1}(q)=& 1-\hat{\Pi}(q)\cr
               =& 1+bg^2\ln(\frac{-q^2}{\mu^2})\cr}
\EQN FullSelfEner$$
We see that the modified propagator has a gauge independent self-energy
and only a trivial gauge dependence originating from the tree part given
by \Eq{GaugeProp}.

It is important to emphasize at this point that the gauge-invariant
self-energies and vertices
obtained by the application of the S-matrix pinch technique do {\sl not}
depend on the particular process employed
 (fermion + fermion $\rightarrow$
fermion + fermion, fermion + fermion $\rightarrow$ gluon + gluon,
 gluon + gluon $\rightarrow$ gluon + gluon, etc.) and are in that
sense universal. This fact can be seen with an explicit calculation, where
one can be convinced that the only
 quantities entering in the definition of the
gauge-independent self-energies and vertices are just the gauge group
structure constants, and that the only difference from process to process
is the external group matrices associated with external-leg wave functions
-due to the different particle assignments- which are, of course, immaterial
to the definition of the things inside.
A very instructive example of an explicit calculation,
where two different processes give rise to {\sl exactly} the same
self-energy
for the $W$ gauge boson, is given in \cite{Degrassi and Sirlin}.

Finally, we conclude this section with a brief presentation of an alternative
formulation of the P.T. introduced in \cite{Degrassi and Sirlin}
in the context of the Standard Model. In this
approach the interaction of gauge bosons with external fermions
is expressed in terms of
current correlation functions, i.e. matrix elements of Fourier transforms
of time-ordered products of current operators
\reference{Another Sirlin}
A.~Sirlin,
\journal Rev. Mod. Phys.; 50,573 (1978)
\endreference
{}.
This is particularly economical because these amplitudes automatically
include several closely related Feynman diagrams. When one of the current
operators is contracted with the appropriate four-momentum, a Ward identity
is triggered. The pinch part is then identified with the contributions
involving the equal-time commutators in the Ward identities, and therefore
involve amplitudes in which the number of current operators has been
decreased by one or more. A basic ingredient in this formulation are the
following equal-time commutators, some of which we will also employ later in
section 3:

$$
\delta(x_0-y_0)[J^{0}_{W}(x),J^{\mu}_{Z}(y)]=
 c^{2}J^{\mu}_{W}(x)\delta^{4}(x-y)
\EQN Commut1$$

$$
\delta(x_0-y_0)[J^{0}_{W}(x),J^{\mu\dagger}_{W}(y)]=
 - J^{\mu}_{3}(x)\delta^{4}(x-y)
\EQN Commut2$$

$$
\delta(x_0-y_0)[J^{0}_{W}(x),J^{\mu}_{\gamma}(y)]=
 J^{\mu}_{W}(x)\delta^{4}(x-y)
\EQN Commut3$$

with $J_{3}^{\mu}\equiv 2(J_{Z}^{\mu}+s^{2}J_{\gamma}^{\mu})$.
 On the other hand
$$
\delta(x_0-y_0)[J^{0}_{V}(x),J^{\mu}_{V^{'}}(y)]= 0
\EQN Commut4$$
where $V,V^{'} \in \{ \gamma,Z \}$.
To demonstrate the method with an example, consider
the vertex $\Gamma_{\mu}$
 shown in Fig 2(b), where now the gauge particles in the loop are W s instead
of
gluons and the incoming and outgoing fermions are massless.
It can be written as follows (with $\xi=1$):
$$
\Gamma_{\mu}=\int \frac{d^{4}k}{{2\pi}^4}
\Gamma_{\mu\alpha\beta}(q,k,-k-q)\int d^{4}x e^{ikx}
<f|T^{*}[J^{\alpha\dagger}_{W}(x)J^{\beta}_{W}(0)]|i>
\EQN Papous$$
When an appropriate momentum, say $k_{\alpha}$,
 from the vertex is pushed into the integral over
$dx$, it gets transformed into a covariant derivative
 $\frac{d}{dx_{\alpha}}$ acting on the time ordered product
$<f|T^{*}[J^{\alpha\dagger}_{W}(x)J^{\beta}_{W}(0)]|i>$.
After using current
conservation and differentiating the
$\theta$-function terms, implicit in the definition of
the $T^{*}$ product, we end up with the left-hand side
 of \Eq{Commut2}.
So, the contribution of each such term is proportional to the
matrix
element of a single current operator,
 namely $<f|J_{3}^{\mu}|i>$; this
is precisely the pinch part. Calling $\Gamma_{\mu}^{P}$
 the total pinch contribution from the
$\Gamma_{\mu}$ of \Eq{Papous}, we find that
$$
\Gamma_{\mu}^{P}= -g^{3}cI_{WW}(Q^2)<f|J_{3}^{\mu}|i>
\EQN PinchPapou$$
where
$$
I_{ij}(q)= i\int (\frac{d^{4}k}{2\pi^{4}})\frac{1}
{(k^{2}-M_{i}^{2})[{(k+q)}^{2}-M_{j}^{2}]}
\EQN IntegralIWW$$
Obviously, the integral in \Eq{IntegralIWW} is the generalization
of the QCD expression \Eq{TotalPinch}
 to the case of massive gauge bosons.
 \section {The gauge invariant three vector boson vertices }
After this brief introduction to the P.T.,
 we now focus on the main topic of
this paper.
In this section we use the S-matrix P.T. to construct at one-loop order
 gauge-independent three-boson vertices
in the context of the Standard Model, when {\sl all} incoming momenta
are off-shell. This problem has first been addressed in the case of
QCD in \cite{Mike and I}. The generalization of the method to the
case of the Standard Model was outlined
in \cite{Degrassi and Sirlin}
and the general structure of the
new vertices was derived;
 however no explicit expressions for the pinch
contributions were reported. Since these contributions are essential for
rendering the final expressions of the static properties of the W bosons
gauge-independent and
infrared finite, we will record in this section their
explicit expressions.

We consider the S-matrix element for the
the process
$e^{+}e^{-} \rightarrow e^{+}\bar{\nu _{e}} e^{-}\nu _{e}$.
Without any loss of generality we will consider all
external fermions to be massless
\reference{footnote}
As explained in \cite{Klassiko}, in general
additional contributions to the gauge independent quantities under
construction arise, when the S- matrix P.T. is applied for external fermions
with different masses, like massive electrons and massless neutrinos.
It is easy to see however that, in the case at hand, all such
contributions do {\sl not}
 kinematically contribute to the definition of the three-vector-boson
vertices, and they will be discarded anyway
\endreference
{}.
The loop diagrams contributing to the S-matrix can be classified
in several distinct classes; vector boson
vertex diagrams (like in Fig.4 and Fig.5)
, vector boson self- energy diagrams (Fig.9)
, fermion-fermion-boson
vertex diagrams (Fig.6(c)),
and box-like diagrams (Figs 6(a),7(a), and 8(a)).
In addition, there are propagator corrections
to the external fermions as well
as disconnected graphs, which are clearly irrelevant for our purposes,
and we therefore omit them.
We can extract a gauge-independent
improper vertex by identifying the part ${\hat{T}}(q,p_{1},p_{2})$
 of the S-matrix
 which is independent of the external
momenta $l_{i}$ and ${\hat{l}}_{i}$, and
 depends only on the momentum
transfers $q, p_{1}, p_{2}$.
The general form of ${\hat{T}}(q,p_{1},p_{2})$ is shown in Fig.3.
It is $\xi$-independent as long as we add
all Feynman graphs {\sl and} parts of Feynman graphs
that depend only on the momentum
transfers $q, p_{1}, p_{2}$. So, to the usual vertex-diagrams
we must
add the vertex-like contributions
extracted from the box-like graphs,
as shown schematically in Fig 6 and Figs 7,8 and their mirror images graphs.
The inclusion of these extra pieces
gives rise to a gauge independent
expression for ${\hat{T}}(q,p_{1},p_{2})$.
Before we record our results a few technical remarks are warranted.
The sum of all contributions mentioned above assumes the form
of an improper vertex, namely
$$
{\hat{T}}(q,p_{1},p_{2}) \equiv {\hat{\Delta}}(q)
{\hat{\Delta}}(p_{1}){\hat{\Delta}}(p_{2}){\hat{\Gamma}}(q,p_{1},p_{2})
\EQN Improper$$
 "sandwiched" between external spinors, not explicitly shown.
 The
propagators ${\hat{\Delta}}$
in \Eq{Improper} are those constructed via the P.T. according to
\cite{Degrassi and Sirlin}; they
have gauge-independent self-energies and only a trivial gauge dependence,
namely that of their tree-level form. This trivial gauge dependence of
all the ${\hat{\Delta}}$ does not appear in ${\hat{T}}(q,p_{1},p_{2})$,
since the external fermions are on-shell and massless.
Therefore, we can recover the gauge-independent proper
 ${\hat{\Gamma}}(q,p_{1},p_{2})$ from ${\hat{T}}(q,p_{1},p_{2})$,
by stripping off the ${\hat{\Delta}}$s,
 as if they had no longitudinal
pieces at all. Another equivallent and more economical way to isolate
the proper vertex
(as described in \cite{Mike and I} and \cite{Degrassi and Sirlin})
is to notice that the gauge boson self-energies can be converted to
gauge-independent ones through P.T.,
up to a missing piece, namely the pinch contribution of the
mirror-graph of Fig 2(e), which clearly is not present in the process
we consider.
 The missing piece may be added by hand
to ${\hat{\Delta}}$, and then
subtracted from ${\hat{\Gamma}}$.
So, the proper vertex emerges if we
neglect all gauge boson self-energy corrections
and subtract instead
half
of the self-energy pinch contribution for each leg.

We now turn to the details of the calculation.
Since the final result, when correctly constructed, is
 {\sl ostensibly} gauge-independent,
we are allowed to perform the calculation in {\sl any} gauge.
We choose the Feynman-t'Hooft gauge ($\xi_{i} = 1$, with
$i= \gamma,Z,W$), since it is certainly the most convenient one.
 As we already explained, in this gauge
 pinching terms arise only from diagrams that contain
elementary three vector boson
 vertices.
 We will use the following notation:
 Scalar propagators are generally denoted as
$$
D_{i}(p) = \frac {1}{p^{2} - M^{2}_{i}}
\EQN prop$$
The trilinear vertex at tree level is given by
$$ \Gamma_{\alpha\beta\gamma}(k_{1},k_{2},k_{3}) =
g_{\alpha\beta}(k_{1}-k_{2})_{\gamma} + g_{\beta\gamma}(k_{2}-k_{3})_{\alpha}
 + g_{\gamma\alpha}(k_{3}-k_{1})_{\beta}
\EQN trive$$
Left and right projectors are defined as :
$$
 P_{R,L}=\frac{1}{2} (1 \pm \gamma _{5})
\EQN P$$
We denote by $(dk) \equiv \frac {d^{4}k}{i(2\pi)^{4}}$ the loop integration
 measure for convergent integrals and by
$(dk) \equiv  \mu ^{4-n}\frac {d^{n}k}{i(2\pi)^{n}}$ the measure for divergent
ones , with $\mu$ the t'Hooft mass scale of dimensional regularization.

To construct the gauge-independent vertex, we add to the regular
vertex graphs of Fig.4 and Fig.5 (which we will call
$V^{i}_{\mu\alpha\beta}$)
the vertex-like pinch parts that arise from the
box diagrams shown in Figs 6b,7b,8c. Clearly, we must also include
the mirror graphs corresponding to Figs 7b,8c
which are hooked on the external fermions of the left-hand side,
and are not shown explicitly.
To begin with,
from the box diagrams of the type shown in Fig 6a ,
only those with two W's in the loop will
 contribute a vertex-like
pinch part.
The rest contain two neutral gauge vector bosons
($\gamma\gamma, \gamma Z, Z\gamma, ZZ$) and, unlike graph 6a,
they also have
crossed graphs, the pinch part of which cancels the pinch part of the
direct diagrams.
This is most easily seen in the framework
of the P.T. formulation of
\cite{Degrassi and Sirlin}; since
the equal-time commutator of the relevant currents
$J_{\gamma},J_{Z}$
is zero (see \Eq{Commut4}, their total pinch contribution vanishes.

The total pinch contribution of diagram 6a
is proportional to $g^{2}\gamma^{\mu} P_{L}$ and can
be written as linear combination of $J_{\gamma}^{\mu}$ and
$J_{Z}^{\mu}$, namely
 $g^{2}\gamma^{\mu} P_{L} = \frac{1}{2}(s^{2}J_{\gamma}^{\mu} +J_{Z}^{\mu})$.
The first term will be alloted to the $\gamma WW$ vertex and the second to the
 ZWW vertex. If we define :
$$
 B_{\mu\alpha\beta}(q,p_{1},p_{2}) =
     \sum_{V} g_{V}^{2}B^{V}_{\mu\alpha\beta}(q,p_{1},p_{2})
\EQN Bup$$
where $B^{V}_{\mu\alpha\beta}$ is the integral :
$$ \eqalign{
B^{i}_{\mu\alpha\beta}(q,p_{1},p_{2}) = &~~~
\int (dk)D_{i}(k)D_{W}(k+p_{1})D_{W}(k-p_{2}) \times \cr
 & \{~g_{\mu\alpha}(2p_{1 \beta}-3k_{\beta})
+g_{\alpha\beta}(k_{\mu}- \frac{3}{2}(p_{1}-p_{2})_{\mu})
-g_{\mu\beta}(2p_{2 \alpha}+3k_{\alpha})~\} \cr}
\EQN Bi$$
and the summation index $V=\gamma ,Z$ refers to the
{\sl internal} $\gamma$ or $Z$ propagator,
then the pinch contribution of these box diagrams is :
$$
(~6b~)_{\gamma WW}
 = (-gs)q^{2}B_{\mu\alpha\beta}(q,p_{1},p_{2})
\EQN 6bPg$$
and
$$
(~6b~)_{ZWW}
 = (-gc)[q^{2}-M^{2}_{Z}] B_{\mu\alpha\beta}(q,p_{1},p_{2})
\EQN 6bPZ$$
for the $\gamma WW$ and ZWW vertices respectively.

 We next look at the pinch contributions of the box diagrams of the type shown
in
 Figs 7 or 8.
There are two such contributions,
depending  on whether the pinching occurs at the side
 of the $W^{+}$ or at the side of the $W^{-}$
 (the latter are shown in Figs 7b,8c).
We call them respectively $B^{+}_{\mu\alpha\beta}$
 and $B^{-}_{\mu\alpha\beta}$ and they
are connected by the relation:
$$
B^{-}_{\mu\alpha\beta}(q,p_{1},p_{2}) = - B^{+}_{\mu\beta\alpha}(q,p_{2},p_{1})
\EQN 12$$
It is clear from Fig.8
that when the neutral
vector boson in the loop is a $Z$,
we have a direct and a crossed graph. These two graphs are different,
since the internal fermion is an $e$ or a $\nu$ respectively.
The pinch parts
of the direct and crossed diagram are again opposite
to each other, but since
the couplings are different, their total sum is not zero, according to
$(\nu We)(eZe)-(\nu Z\nu)(\nu We) = -g_{Z}(\nu We)$

Then $B^{+}_{\mu\alpha\beta} $ is given by :
$$
 B^{+}_{\mu\alpha\beta}(q,p_{1},p_{2})
=\sum _{V} g_{V}^{2}G_{\mu\alpha\beta}^{V}(q,p_{1},p_{2})
\EQN Bplus$$
where $G_{\mu\alpha\beta}^{V}$ is the following integral :
$$\eqalign{
G_{\mu\alpha\beta}^{V}(q,p_{1},p_{2})  =&~~~~
     \int (dk)D_{V}(k)D_{W}(k+p_{1})D_{W}(k-p_{2})\times \cr
&[~g_{\alpha\beta}~(3k+2p_{1}-3p_{2})_{\mu}
 + g_{\beta\mu}~(-k+4p_{2})_{\alpha} + g_{\alpha\mu}~(3k-2p_{1})_{\beta}~]
  \cr}
\EQN GV$$
and $V$ is again summed over the internal $\gamma$ and $Z$ propagator.
  Finally, the pinch parts of the box diagrams of this type  are :
$$
(7b) + (8c) =
-g_{V}~(p_{2}^{2}-M_{W}^{2})B^{-}_{\mu\alpha\beta}
\EQN 7b8c$$
$$
(mirror~ image~ of ~ 7b) + (mirror~ image~ of ~ 8c) =
-g_{V}~(p_{1}^{2}-M_{W}^{2})B^{+}_{\mu\alpha\beta}
\EQN mirror7b8c$$
where $g_{V}$ is equal to $gs$ or $gs$ depending on which
gauge-independent vertex
($\gamma W^{+}W^{-}$ or $ZW^{+}W^{-}$) we consider.
Notice the presence
of the typical inverse-propagator-like factor
 $D_{W}^{-1}(p_{1})= p_{1}^{2}-M_{W}^{2}$, which always multiplies
expressions originating
 from pinching.
Before we proceed, we record the result of $q^{\mu}$ acting on
$G_{\mu\alpha\beta}^{V}$ of \Eq{GV}, which will be used in the next section:
$$\eqalign{
q^{\mu}G_{\mu\alpha\beta}^{V} =
 &~~2g_{\alpha\beta}[I_{VW}(p_{1})-I_{VW}(p_{2})] \cr
&+\int (dk)D_{V}(k)D_{W}(k+p_{1})D_{W}(k-p_{2})
[~q_{\alpha}k_{\beta} + q^{\rho}\Gamma_{\rho\alpha\beta}(-p_{2}+k,-k,p_{2})~]
  \cr}
\EQN qG$$
where $I_{ij}(p)$ has been defined in \Eq{IntegralIWW}.

As we already explained at the beginning of this section, in  order
to isolate the proper vertex we must subtract half of the
contribution of the self-energy pinch graphs for
the propagator of each leg.
All such contributions are of the general form
$$
\Gamma_{\mu\alpha\beta}(q,p_{1},p_{2})I_{JW}(p)
\EQN Lambda$$
where $J=W,\gamma ,Z$~ and
 $p=q,p_{1},p_{2},$ .$\Gamma_{\mu\alpha\beta}$ is the tree-level
$W^{+}W^{-}$ or $ZW^{+}W^{-}$ vertex (Fig. 6(d)). We will not reproduce
the details of this last step here.
(see \cite{Degrassi and Sirlin} for more details)

Finally, the one loop gauge-independent trilinear gauge boson vertices
( after we pull out a factor of $-igc$ or $-igs$) are
$$\eqalign{
\hat{\Gamma}_{\mu\alpha\beta} &=~
V_{\mu\alpha\beta}
- (q^{2}-M^{2}_{V})B_{\mu\alpha\beta}
- (p_{1}^{2}-M_{W}^{2})B^{+}_{\mu\alpha\beta}
- (p_{1}^{2}-M_{W}^{2})B^{-}_{\mu\alpha\beta} \cr
&-2g^{2}\Gamma_{\mu\alpha\beta}
[I_{WW}(q)+s^{2}I_{\gamma W}(p_{1})+c^{2}I_{ZW}(p_{1})
+s^{2}I_{\gamma W}(p_{2})+c^{2}I_{ZW}(p_{2})] \cr}
\EQN GIV$$
 where
$V_{\mu\alpha\beta}=\sum_{i}V^{i}_{\mu\alpha\beta}$
are the usual one loop corrections to the
$\gamma W^{+}W^{-}$ or $ZW^{+}W^{-}$ vertex
in the Feynman gauge ( $\xi_{i}=1$ , $i=\gamma ,Z,W$ ).
The one loop diagrams of the $\gamma WW$ vertex in this gauge
are shown in Figs 4,5. There are a few additional graphs for
the $ZW^{+}W^{-}$ vertex, due to the tree-level coupling of $Z$ to
the Higgs.

The vertex constructed via the S-matrix P.T. in the way we described above
represents the most general case, since all three incoming momenta
are off-shell. If one is interested in the simpler case, where the two
incoming momenta of $W^{+}$ and $W^{-}$ are on shell, and only the photon
momentum is off-shell, as is the case, for example, in the process
$e^{+}e^{-} \rightarrow W^{+}W^{-}$, \Eq{GIV} reduces to the following
expression:
$$
\hat{\Gamma}_{\mu\alpha\beta}|_{p_{1}^{2}=p_{2}^{2}=M^{2}_{W}} =
V_{\mu\alpha\beta}
- (q^{2}-M^{2}_{V})B_{\mu\alpha\beta}
- 2g^{2}\Gamma_{\mu\alpha\beta}I_{WW}(q)
\EQN OnShellGIV$$

This is of course the same answer one obtains by applying the P.T.
{\sl directly} to the S-matrix of $e^{+}e^{-} \rightarrow W^{+}W^{-}$.
Clearly, in that case the only vertex-like contribution comes from
graph Fig. 6(a), and is just $g_{V}B_{\mu\alpha\beta}q^{2}$, whereas the term
$-2g_{V}g^{2}\Gamma_{\mu\alpha\beta}I_{WW}(q)$
arises exactly as before,
namely as a leftover from the construction of
gauge independent $\gamma\gamma$, $\gamma Z$, and $ZZ$ self-energies.
We note
that with the two W's on shell the $B_{\mu\alpha\beta}$ function becomes
infrared divergent. This divergence, however, cancels against the infrared
 divergence that the $V^{1}_{\mu\alpha\beta}$ diagram develops as well,
when the
two W's are considered on shell. This will be shown in section 5.

\section {The Ward identity for the $\gamma$WW vertex}

 In this section we show that the one-loop gauge-invariant $\gamma$WW vertex
$\hat{\Gamma}_{\mu\alpha\beta}$  constructed in the previous section via the
 S-matrix P.T. satisfies a simple
 QED-like Ward identity, which relates it to
the gauge invariant one-loop self-energy
$\hat{\Pi}_{\alpha\beta}$
 of the W, constructed in \cite{Degrassi and Sirlin} (Eq.(19)).

The proof of the Ward identity is rather lengthy
and technically involved; we therefore
start this section by recording the result,
 for the reader who wants a quick tour through the paper.
The gauge-invariant $ \gamma W^{+} W^{-}$ vertex
${\hat{\Gamma}}_{\mu\alpha\beta}(q,p_{1},p_{2})$
and the gauge-invariant W self-energy
$\hat{\Pi} _{\alpha\beta}$ satisfy the following Ward identity:

$$
q^{\mu}{\hat{\Gamma}}_{\mu\alpha\beta}(q,p_{1},p_{2}) =
     {\hat{\Pi}}_{\alpha\beta}(p_{1}^{2})~ - ~
        {\hat{\Pi}}_{\alpha\beta}(p_{2}^{2})
\EQN WI$$

with
$$
\hat{\Pi}_{\alpha\beta}(p^{2}) = \Pi_{\alpha\beta}(p^{2})
   -4g^{2}(p^{2}-M^{2}_{W})[c^{2}I_{ZW}(p^{2})+s^{2}I_{\gamma W}(p^{2})]
\EQN Pi$$
where $\Pi_{\alpha\beta}(p)$ are the usual one loop  W self energy
contributions with tadpole contributions incorporated , in the Feynman
gauge.
The second term is the pinch part , which is given diagramaticaly in
Fig. 9 (pinch).

We now proceed to prove \Eq{WI}.
In doing so, we find it more economical to act with $q_{\mu}$
{\sl directly}
on the Feynman graphs,
instead of first evaluating them and then act with $q_{\mu}$ on the
 final answer.
  In this way,
 cancellations among entire graphs become immediately apparent.
 The W self-energy diagrams that
are relevant for the construction of the R.H.S. of \Eq{WI}
are shown in Fig 9. Seagull and tadpole diagrams are omitted,
since they do not depend on
the momenta $p_{1}$ or  $p_{2}$,
 and thus cancel when we form
the difference of the two self-energies in \Eq{WI}.

We now contract the R.H.S. of \Eq{GIV} with $q^{\mu}$.
The following identity is frequently used
$$\eqalign{
k_{2}^{\beta} \Gamma_{\alpha\beta\gamma}(k_{1},k_{2},k_{3}) &=
 P_{\alpha\gamma}(k_{1}) - P_{\alpha\gamma}(k_{3})\cr
 &= g_{\alpha\gamma}[D_{i}^{-1}(k_{1})-D_{j}^{-1}(k_{3})] -
k_{1\alpha}k_{1\gamma} + k_{3\alpha}k_{3\gamma}
 + g_{\alpha\gamma}\Delta M_{ij}^{2}\cr}
\EQN ide$$
with ~~$ \Delta M_{ij}^{2} = M_{i}^{2} - M_{j}^{2} $

We start with the
vertex pinch parts $B_{\mu\alpha\beta}^{+}, B_{\mu\alpha\beta}^{-}$ and
$B_{\mu\alpha\beta}$; using \Eq{qG} we obtain :
$$\eqalign{
-q^{\mu}B_{\mu\alpha\beta}^{+}D_{W}^{-1}(p_{1}) &=
{}~~~- D_{W}^{-1}(p_{1})\sum_{V} g^{2}_{V}
\Biggl\lbrack 2g_{\alpha\beta}[I_{VW}(p_{1})-I_{VW}(p_{2})] \cr
&+\int (dk)D_{V}(k)D_{W}(k+p_{1})D_{W}(k-p_{2})
[~q_{\alpha}k_{\beta} + q^{\rho}\Gamma_{\rho\alpha\beta}(-p_{2}+k,-k,p_{2})~]
\Biggr\rbrack \cr}
\EQN qBplus$$
and
$$\eqalign{
-q^{\mu}B_{\mu\alpha\beta}^{-}D_{W}^{-1}(p_{2}) &=
- D_{W}^{-1}(p_{2})\sum_{V} g^{2}_{V}
\Biggl\lbrack 2g_{\alpha\beta}[I_{VW}(p_{1})-I_{VW}(p_{2})] \cr
&+\int (dk)D_{V}(k)D_{W}(k+p_{1})D_{W}(k-p_{2})
[~q_{\beta}k_{\alpha} + q^{\rho}\Gamma_{\rho\alpha\beta}(-p_{1}-k,p_{1},k)~]
\Biggr\rbrack \cr}
\EQN qBminus$$
On the other hand
$q^{\mu}B_{\mu\alpha\beta}=0$, since the
$J_{\gamma}^{\mu}$ current is divergence-less.

We continue with the self-energy pinch parts of the vertex;
using \Eq{ide}, we get :
$$
-2g^{2}g_{\alpha\beta}
[D_{W}^{-1}(p_{2})-D_{W}^{-1}(p_{1})]
[I_{WW}(q^{2})+s^{2}I_{\gamma W}(p_{1}^{2})+c^{2}I_{ZW}(p_{1}^{2})
+s^{2}I_{\gamma W}(p_{2}^{2})+c^{2}I_{ZW}(p_{2}^{2})]
\EQN qSE$$
where terms proportional to $p_{1\alpha}$ or $p_{2\beta}$
, that are zero
on shell, have been discarded \cite{african shield}.

 The first terms of \Eq{qBplus} and \Eq{qBminus} when combined with the last
 four terms of \Eq{qSE} give
$$
\sum_{V} 2g_{V}^{2}g_{\alpha\beta} \times
$$
$$
[~(D_{W}^{-1}(p_{1})+D_{W}^{-1}(p_{2}))(I_{VW}(p_{1})-I_{VW}(p_{2})
- (D_{W}^{-1}(p_{2})-D_{W}^{-1}(p_{1}))(I_{VW}(p_{1})+I_{VW}(p_{2})~]
$$
$$
=\sum_{V} 4g_{V}^{2}g_{\alpha\beta}
(~D_{W}^{-1}(p_{1})I_{VW}(p_{1})-D_{W}^{-1})(p_{2})I_{VW}(p_{2})~)
\EQN PSE $$
We recognize in this last term the difference of the pinch contributions
 that render
the W self-energies, defined at $p_{1}$ and $p_{2}$, gauge independent.

We continue by calculating the divergence of the diagrams of Fig 4(1,2)
$V^{1}_{\mu\alpha\beta}$  and $V^{2}_{\mu\alpha\beta}$, given by
$$\eqalign{ V^{1,2} =
-g^{2}_{V}\int & (dk)D_{V}(k)D_{W}(k+p_{1})D_{W}(k-p_{2})\times \cr
&\times \Gamma_{\mu\rho\sigma}(q,p_{1}+k,p_{2}-k)
\Gamma_{\alpha\lambda}^{~~\rho}(p_{1},k,-p_{1}-k)
\Gamma_{\beta}^{~\sigma\lambda}(p_{2},k-p_{2},-k)\cr}
\EQN d(1,2)$$
When contracting the expression above with $q^{\mu}$ the identity \Eq{ide} is
 triggered and we get :
$$ \eqalign{
q^{\mu}V_{\mu\alpha\beta}^{1,2} = -&g^{2}_{V}\int
(dk) D_{V}(k)D_{W}(k+p_{1})D_{W}(k-p_{2}) \times \cr
&\Gamma_{\alpha\lambda}^{~~\rho}(p_{1},k,-p_{1}-k)
\Gamma_{\beta}^{~\sigma\lambda}(p_{2},k-p_{2},-k)\times \cr
 &[ g_{\rho\sigma}(D_{W}^{-1}(p_{2}-k)-D_{W}^{-1}(p_{1}-k)) +
 (p_{1}+k)_{\rho}(p_{1}+k)_{\sigma} - (p_{2}-k)_{\rho}(p_{2}-k)_{\sigma} ]\cr}
\EQN step1$$

 We first concentrate on the $g_{\rho\sigma}$ term. It reads :
$$\eqalign{
-g^{2}_{V} \int(dk)& [ D_{V}(k)D_{W}(k+p_{1}) -
D_{V}(k)D_{W}(k-p_{2})] \times \cr
&\Gamma_{\alpha\lambda\sigma}(p_{1},k,-p_{1}-k)
\Gamma_{\beta}^{~\sigma\lambda}(p_{2},k-p_{2},-k) \cr }
\EQN gterm$$
It contains only two internal propagators
 and could be identified with the self-energy
graphs $\Pi ^{1,2}_{\alpha\beta}$ of Fig 9(1,2) {\sl if} the
$\Gamma_{\alpha\lambda\sigma}\Gamma_{~~\beta}^{\sigma\lambda}$
factor had the appropriate momenta arguments.
To convert this terms to the desired form,
we use $p_{1}+p_{2}+q = 0$ to write
$$
\Gamma_{~~\beta}^{\sigma\lambda}(k-p_{2},-k,p_{2}) =
\Gamma_{~~\beta}^{\sigma\lambda}(k+p_{1},-k,-p_{1}) +
2q^{\lambda}g^{\sigma}_{\beta} - q^{\sigma}g_{~\beta}^{\lambda} -
q_{\beta}g^{\sigma\lambda}
\EQN G1$$
for the first part, and
$$
\Gamma_{\alpha\lambda\sigma}(p_{1},k,-p_{1}-k) =
\Gamma_{\alpha\lambda\sigma}(p_{2}-k,-p_{2},k) +
2q_{\lambda}g^{\rho}_{\alpha} - q^{\rho}g_{\alpha\lambda} - q_{\alpha}g^{\rho}
_{\lambda}
\EQN G2$$
for the second part.
After bringing the $\Gamma$s in the correct form,
we recognize that the the terms left-over in \Eq{G1} and \Eq{G2}
 are equal to
$-\sum_{i} q^{\mu}V^{i}_{\mu\alpha\beta}$
 for $i=21,22,23,24$, namely the
negative of the divergence of the diagrams in Fig 5 (21,22,23,24);
thus all these contributions will cancel in the L.H.S. of \Eq{WI}.

So, the term proportional to
$g_{\rho\sigma}$ is equal to:
$$
-(\Pi^{1,2}_{\alpha\beta}(p_{1})- \Pi^{1,2}_{\alpha\beta}(p_{2}))
 - q^{\mu}V_{\mu\alpha\beta}^{21,20} - q^{\mu}V_{\mu\alpha\beta}^{23,22}
\EQN gtermfin$$

We next examine the term  $(p_{1}+k)_{\rho}(p_{1}+k)_{\sigma} -
(p_{2}-k)_{\rho}(p_{2}-k)_{\sigma}$ of \Eq{step1}. Contracting each momentum
of this term with the appropriate vertex
in order to exploit \Eq{ide}, we obtain
$$
g^{2}_{V} \int
(dk) D_{V}(k)D_{W}(k+p_{1})D_{W}(k-p_{2}) \times
$$
$$\eqalign{
\Biggl\lbrack
& q^{\rho}\Gamma_{\rho\lambda\beta}(k-p_{2},-k,p_{2})[~(~ D_{W}^{-1}(p_{1})
 - D_{V}^{-1}(k)~)g_{~\alpha}^{\lambda}+\Delta M_{WV}^{2}g_{~\alpha}^{\lambda}+
k_{\alpha}k^{\lambda}~] \cr
+~ & q^{\rho}\Gamma_{\alpha\lambda\rho}(p_{1},k,-p_{1}-k)[~(D_{W}^{-1}(p_{2}
-D^{-1}_{V}(k)~)g^{\lambda}_{\beta}+\Delta M^{2}_{WV}g^{\lambda}_{\beta} +
k_{\beta}k^{\lambda}~]\Biggr\rbrack \cr}
\EQN Lterm$$
where we have omitted terms proportional to $p_{1\alpha}$ or
$p_{2\beta} $ that are zero on shell.
Using the identity :
$$
- eg_{V}^{2} \Delta M^{2}_{WV} ~=~
 -b_{V} g^{3} sin^{3} \theta M^{2}_{W}
\EQN Gold$$
where $b_{\gamma}=+1 $ and $ b_{Z}=-1$,
 we recognize that the
terms of \Eq{Lterm} proportional to $\Delta M^{2}_{WV}$
are equal to $-\sum_{i} q^{\mu}V_{\mu\alpha\beta}^{i}$
of Fig 4(3,4,5,6); all these contributions will also cancel
in \Eq{WI}.

We continue
by noticing that in the
$D_{V}^{-1}(k)$ terms of \Eq{Lterm} the $V=\gamma,Z$ propagators cancel,
and the resulting expression depend on V only
through the coupling $g_{V}^{2}$, namely
$$
-g^{2}_{V} \int
(dk) D_{W}(k+p_{1})D_{W}(k-p_{2})q^{\rho}\Gamma_{\rho\alpha\beta}(q,p_{1}-k,
p_{2}+k)
\EQN DVterm$$
So, when we add the diagrams of Fig 4(1) and Fig 4(2)
 the couplings will give $-g^{2}$,
and after using \Eq{ide} in \Eq{DVterm},
 and shifting the integration variables,
  we get for this last term
$$\eqalign{
&=g^{2}
 \int (dk) D_{W}(p_{1}+k)D_{W}(p_{2}-k)~
(4q \cdot k)~g_{\alpha\beta} \cr
&=2g^{2}g_{\alpha\beta}(~D_{W}^{-1}(p_{2})-D_{W}^{-1}(p_{1})~)I_{WW}(q)
\cr}
\EQN DVfin$$
which will cancel against the first term of \Eq{qSE}.

For the $k_{\alpha}k^{\lambda}$ and $k_{\beta}k^{\lambda}$ terms of \Eq{Lterm}
we perform the contractions $k^{\lambda}\Gamma_{\rho\lambda\beta}$ and
$k^{\lambda}\Gamma_{\rho\alpha\lambda}$, and we get :
$$
= g_{V}^{2} \int (dk)D_{V}(k)D_{W}(k+p_{1})D_{W}(k-p_{2}) \times
$$
$$
[~k_{\alpha}q_{\beta}D_{W}^{-1}(p_{2}) + k_{\beta}q_{\alpha}D_{W}^{-1}(p_{1})
 + q\cdot (k-p_{2})k_{\alpha}k_{\beta} + q\cdot (k+p_{1})k_{\alpha}k_{\beta}~]
\EQN kkterm$$
and terms that are zero on shell have been omitted. Collecting the terms
proportional to
 $D_{W}^{-1}(p_{1})$ and $D_{W}^{-1}(p_{2})$ from the above equation and
 from \Eq{Lterm}, we immediately see that they cancel
against the second terms of
\Eq{qBplus} and \Eq{qBminus}.
It is now important to recognize that the two remaining terms
of \Eq{kkterm} are equal to the divergence of the vertex
diagrams containing a ghost loop [Fig 4, diagrams(10,11,12,13)],
and their presence
is crucial for recovering
the W self energies on the R.H.S. of \Eq{WI}.
Indeed, from the vertex graphs with ghosts we obtain
$$
q^{\mu}(V^{10}_{\mu\alpha\beta}+V^{12}_{\mu\alpha\beta})
= \Pi^{8}_{\alpha\beta}(p_{1})+\Pi^{8}_{\alpha\beta}(p_{2})
\EQN g1$$
and
$$
q^{\mu}(V^{11}_{\mu\alpha\beta}+V^{13}_{\mu\alpha\beta})
= \Pi^{9}_{\alpha\beta}(p_{1})-\Pi^{9}_{\alpha\beta}(p_{2})
\EQN g2$$
where the diagrams
$$
\Pi^{9}_{\alpha\beta}=\Pi^{11}_{\alpha\beta} ~~ , ~~
\Pi^{8}_{\alpha\beta}=\Pi^{10}_{\alpha\beta}
\EQN WSEG$$
are the ghost one loop corrections to the W self-energy,
and are shown in Fig. 9 (8,9,10,11). Evidently, the vertex
ghost diagrams
contribute {\sl only} half of the necessary
self-energy ghost diagrams,
while the other half will be provided by the last two terms of
\Eq{kkterm}.

 In a straightforward way we also arrive to the following results :

 Higgs diagrams
$$
 q^{\mu}V^{14}_{\mu\alpha\beta} =
 \Pi^{3}_{\alpha\beta}(p_{1})-\Pi^{3}_{\alpha\beta}(p_{2})
-q^{\mu}V^{15}_{\mu\alpha\beta}-q^{\mu}V^{16}_{\mu\alpha\beta}
\EQN Higgs1$$
$$
 q^{\mu}V^{17}_{\mu\alpha\beta} =
 \Pi^{6}_{\alpha\beta}(p_{1})-\Pi^{6}_{\alpha\beta}(p_{2})
\EQN Higgs2$$

 Goldstone boson diagrams
$$
 q^{\mu}V^{7}_{\mu\alpha\beta} =
 \Pi^{5}_{\alpha\beta}(p_{1})-\Pi^{5}_{\alpha\beta}(p_{2})
\EQN Gold1$$
$$
 q^{\mu}V^{8}_{\mu\alpha\beta} =
 \Pi^{4}_{\alpha\beta}(p_{1})-\Pi^{4}_{\alpha\beta}(p_{2})
\EQN Gold2$$
$$
 q^{\mu}V^{9}_{\mu\alpha\beta} =
 \Pi^{7}_{\alpha\beta}(p_{1})-\Pi^{7}_{\alpha\beta}(p_{2})
\EQN Gold3$$

For the fermion diagrams the Ward identity is trivially satisfied
in a QED-like fashion. For the rest of the vertex diagrams that have not
been treated thus far,
 the result of their contraction with $q^{\mu}$ gives
zero on shell. Adding all relevant equations together we arrive at
the advertised
Ward identity of \Eq{WI}, a major result of this paper.
It is important to notice that the pinch contributions have been
instrumental for the validity of \Eq{WI}.

\section {Magnetic dipole and electric quadruple moments of the W}

In the last two sections we constructed the gauge-independent
 $\gamma W^{+}W^{-}$ vertex and we proved the Ward identity it satisfies.
In this section we will proceed to extract from this vertex its
 contributions to the magnetic dipole $\mu_{W}$ and electric quadrupole
$Q_{W}$. Following the parametrization of
\Eq{dipole} and \Eq{quad}, we need to determine the
quantities ${\Delta\kappa}_{\gamma}$ and ${\Delta Q}_{\gamma}^{P}$,
or equivalently $\kappa_{\gamma}$ and $\lambda_{\gamma}$.
Since they originate from the new $\gamma W^{+}W^{-}$
vertex, unlike
 the expressions recorded in \cite{Lahanas},
 they will be gauge independent and infrared finite.
 We use "hats" to indicate these
new {\sl gauge-independent} and {\sl infrared finite}
 contributions, namely
${\hat{\Delta}\kappa}_{\gamma}$
and  ${\hat{\Delta} Q}_{\gamma}$, respectively.

As we explained in section 3, the new gauge invariant vertex is build up
from the usual vertex diagrams calculated in the Feynman-t'Hooft gauge
 and the vertex-like pinch contributions we extract from
box diagrams. So, if we denote by
${\Delta\kappa}_{\gamma}^{(\xi=1)}$
and  ${\Delta Q}_{\gamma}^{(\xi=1)}$, respectively, the contributions
of the usual vertex diagrams ($\xi=1$), and by
${\Delta\kappa}_{\gamma}^{P}$
and  ${\Delta Q}_{\gamma}^{P}$ the analogous contributions
of the pinch parts, we clearly have the following relations:
$$
{\hat{\Delta}\kappa}_{\gamma}= {\Delta\kappa}_{\gamma}^{(\xi=1)}
+ {\Delta\kappa}_{\gamma}^{P}
\EQN Finalkappa$$
and
$$
{\hat{\Delta} Q}_{\gamma} = {\Delta Q}_{\gamma}^{(\xi=1)}
 + {\Delta Q}_{\gamma}^{P}
\EQN FinalQ$$

The task of actually calculating
${\hat{\Delta}\kappa}_{\gamma}$ and ${\hat{\Delta} Q}_{\gamma}$ is greatly
facilitated by the fact that the quantities
${\Delta\kappa}_{\gamma}^{(\xi=1)}$ and ${\Delta Q}_{\gamma}^{(\xi=1)}$
have already been calculated in \cite{Lahanas}. It must be emphasized
however that the expression for ${\Delta\kappa}_{\gamma}^{(\xi=1)}$
(but not  ${\Delta Q}_{\gamma}^{(\xi=1)}$)
is infrared {\sl divergent} for $Q^2 \not = 0$ due to the presence of the
following double integral over the Feynman parameters (t,a), given in
Eq.(26) of \cite{Lahanas}:

$$\eqalign{
R &= -(\frac{\alpha_{\gamma}}{\pi})\frac{Q^{2}}{M^{2}_{W}}
\int^{1}_{0} da \int^{1}_{0}
\frac{dtt}{t^{2}-t^{2}(1-a)a(\frac{4Q^2}{M^2_{W}})}\cr
&= -(\frac{\alpha}{2\pi})\frac{Q^{2}}{M^{2}_{W}}
\int^{1}_{0}
 \frac{da}{1-(1-a)a(\frac{4Q^2}{M^2_{W}}}) \int^{1}_{0}
\frac{dx}{x}\cr}
\EQN BadGuy$$
which originates from
the graph of Fig.4 (1,2), when one of the internal propagators
is a virtual photon.
($\alpha_{\gamma} \equiv \frac{e^{2}}{4\pi}$ is the fine structure constant).

We only need therefore to determine the expressions for
 ${\Delta Q}_{\gamma}^{P}$ and ${\Delta\kappa}_{\gamma}^{P}$;
this is equivalent to determining the pinch contributions to the
functions $a_{1}(Q^{2})$, $a_{2}(Q^{2})$ and $a_{3}(Q^{2})$
defined in \Eq{1loopParametrization}, which we will call
$a_{1}^{P}(Q^{2})$ $a_{2}^{P}(Q^{2})$ and $a_{3}^{P}(Q^{2})$.
For on-shell $W^{+}$ and $W^{-}$ pinch contribution originate
only from the $B_{\mu\alpha\beta}^{V}$ term in \Eq{OnShellGIV}.
To establish contact with \cite{Lahanas}, we use the following
identity to parameterize the denominator of the integrals:
$$
\frac{1}{ABC}= \int_{0}^{1}da \int_{0}^{1} dt
\frac{2t}{{\{[A(1-a)+Ba]t + C(1-t)]\}}^{3}}
\EQN FeynmanParam$$

The momentum integration can be immediately performed
(We remind the reader that $B_{\mu\alpha\beta}$ is ultraviolet finite,
so no regularization is needed).
In the limit of interest, namely $p_{1}^{2}=p_{2}^{2}=M_{W}^{2}$,
we find:
$$
B_{\mu\alpha\beta}^{V}= -\frac{1}{2}\frac{\alpha_{V}}{4\pi}
\frac{q^{2}}{M^{2}_{W}}
\int_{0}^{1}da \int_{0}^{1}(2tdt) \frac{F_{\mu\alpha\beta}}{L_{V}^{2}}
\EQN ParamB$$
with
$\alpha_{V}=\frac{g^{2}_{V}}{4\pi}$ and

$$\eqalign{
F_{\mu\alpha\beta}&=(\frac{3}{2} + at)g_{\alpha\beta}{(p_1-p_2)}_{\mu}
 + (3at+2)g_{\beta\mu}{p_{2}}_{\alpha} -
 (3at+2)g_{\beta\mu}{p_{1}}_{\beta}\cr
&= 2(\frac{3}{2} + at)g_{\alpha\beta}{\Delta}_{\mu}+
2(3at+2)[g_{\alpha\mu}Q_{\beta}-g_{\beta\mu}Q_{\alpha}]\cr}
\EQN DefineF$$
and
$$\eqalign{
L^{2}_{V}&= t^{2}-t^{2}a(1-a)
(\frac{q^{2}}{M^{2}_{W}}) + (1-t)\frac{M^{2}_{V}}{M^{2}_{W}}\cr
&= t^{2}-t^{2}a(1-a)(\frac{4Q^{2}}{M^{2}_{W}}) +
 (1-t)\frac{M^{2}_{V}}{M^{2}_{W}}\cr}
\EQN DefineL$$

from which immediately follows that
$$
a_{1}^{P}(Q^{2})= -\frac{1}{2}\frac{Q^{2}}{M^{2}_{W}}
\sum_{V}\frac{\alpha_{V}}{\pi} \int_{0}^{1}da \int_{0}^{1}(2tdt)
\frac{2(\frac{3}{2} + at)}{L^{2}_{V}}
\EQN alpha1P$$
,
$$
a_{2}^{P}(Q^{2})= -\frac{1}{2} \frac{Q^{2}}{M^{2}_{W}}
\sum_{V} \frac{\alpha_{V}}{\pi}
 \int_{0}^{1}da \int_{0}^{1}(2tdt)
\frac{2(2+3at)}{L^{2}_{V}}
\EQN alpha2P$$
and since there is no term proportional to $\Delta_{\mu}Q_{\alpha}Q_{\beta}$
in \Eq{DefineF},
$$
a_{3}^{P}(Q^{2})=0
\EQN alpha3P$$

So, using \Eq{1loopdeltakappa} and \Eq{1loopdeltaQ} we have for
${\Delta\kappa}_{\gamma}^{P}$ and ${\Delta Q}_{\gamma}^{P}$:
$$
{\Delta\kappa}_{\gamma}^{P}= -\frac{1}{2} \frac{Q^{2}}{M^{2}_{W}}
\sum_{V} \frac{\alpha_{V}}{\pi} \int_{0}^{1}da \int_{0}^{1}(2tdt)
\frac{(at-1)}{L^{2}_{V}}
\EQN ligoulaki$$
and
$$
{\Delta Q}_{\gamma}^{P} = 0
\EQN tipota$$

It is important to notice that even though ${\Delta Q}_{\gamma}^{P} = 0$
{\sl both} $\mu_{W}$ and $Q_{W}$ will assume values different
than those predicted in the
$\xi=1$ gauge. That this is so may be seen from \Eq{deltaQ}
 and \Eq{deltakappa}; clearly,
even though the value of $\lambda_{\gamma}$ does not change, the value of
$\kappa_{\gamma}$ changes, and this change affects both
$\mu_{W}$ and $Q_{W}$ through \Eq{dipole} and \Eq{quad}.
In the expression given in \Eq{ligoulaki}
the first term (for V=Z) is infrared finite (since $M_{Z} \not = 0$),
whereas the second term (for $V=\gamma$) is infrared divergent, since
$M_{\gamma}=0$. Calling this second term $\Theta$ we have
$$
\Theta= -\frac{1}{2}
(\frac{\alpha_{\gamma}}{\pi})\frac{Q^{2}}{M^{2}_{W}}
 \int_{0}^{1}da \int_{0}^{1}dt
\frac{2t(at-1)}{t^{2}[1-a(1-a)\frac{4Q^{2}}{M^{2}_{W}}]}
\EQN Theta$$
which can be rewritten as
$$
\Theta=  -R - (\frac{\alpha_{\gamma}}{\pi})\frac{Q^{2}}{M^{2}_{W}}
\int_{0}^{1}da\frac{a}{1-a(1-a)\frac{4Q^{2}}{M^{2}_{W}}}
\EQN AlmostThere$$
where R is the infrared divergent integral
defined in \Eq{BadGuy}. On the other hand,
the second term in \Eq{AlmostThere}
is infrared finite.
Clearly, including the first term of \Eq{AlmostThere} in the value of
${\hat{\Delta}\kappa}_{\gamma}$ {\sl exactly} cancels the infrared divergent
contribution of \Eq{BadGuy}, thus giving rise to an infrared finite expression
for ${\hat{\Delta}\kappa}_{\gamma}$.
So, after the infrared divergent part of \Eq{Theta} is cancelled,
${\Delta\kappa}_{\gamma}^{P}$ is given by the following expression:
$$
{\Delta\kappa}_{\gamma}^{P}= \Theta_{\gamma} + \Theta_{Z}
\EQN AntePali$$
with $\Theta_{\gamma}$ the second term in \Eq{AlmostThere}, and $\Theta_{Z}$
the second term in \Eq{ligoulaki}, namely
$$
\Theta_{\gamma}=
- (\frac{\alpha_{\gamma}}{\pi})\frac{Q^{2}}{M^{2}_{W}}
\int_{0}^{1}da\frac{a}{1-a(1-a)\frac{4Q^{2}}{M^{2}_{W}}}
\EQN ThetaGamma$$
and
$$
\Theta_{Z}=
 -\frac{Q^{2}}{M^{2}_{W}}
 (\frac{\alpha_{Z}}{\pi}) \int_{0}^{1}da \int_{0}^{1}dt
\frac{t(at-1)}{L^{2}_{Z}}
\EQN ThetaZeta$$
and from \Eq{Finalkappa}
$$
{\hat{\Delta}\kappa}_{\gamma} = {[{\Delta\kappa}_{\gamma}^{(\xi=1)}]}_{if} +
\Theta_{\gamma} + \Theta_{Z}
\EQN End$$
where the subscript $(if)$ in the first term of the R.H.S. indicates
that the contribution from the $\xi=1$ gauge is now genuinely
infrared finite (the authors of \cite{Lahanas} removed the infrared
divergent contribution by hand).
Finally, the magnetic dipole moment $\mu_{W}$ and electric quadrupole
moment $Q_{W}$ are given by
$$
\mu_{W}= \frac{e}{2M_{W}}(2+{\hat{\Delta}\kappa}_{\gamma})
\EQN Finaldipole$$
and
$$
Q_{W}= -\frac{e}{M^{2}_{W}}(1+{\hat{\Delta}\kappa}_{\gamma}
+ 2{\hat{\Delta} Q}_{\gamma})
\EQN Finalquad$$
Both ${\Delta Q}_{\gamma}^{(\xi=1)}$
 and ${\Delta\kappa}_{\gamma}^{(\xi=1)}$
have been computed numerically in \cite{Lahanas}.
We now proceed to compute the integrals in \Eq{ThetaGamma}
 and \Eq{ThetaZeta}, which
determine ${\Delta\kappa}_{\gamma}^{P}$.
It is elementary to evaluate $\Theta_{\gamma}$.
 Setting
$\Theta_{\gamma}=-(\frac{\alpha_{\gamma}}{\pi}){\hat{\Theta}}_{\gamma}$
we have:
$$\eqalign{
{\hat{\Theta}}_{\gamma}&= \frac{2}{\Delta}
[arctg(\frac{1}{\Delta})- arctg(\frac{-1}{\Delta})],
{}~~for~~ Q^{2}<M_{W}^{2}\cr
&= -4 , ~~ for~~ Q^{2}=M_{W}^{2}\cr
&= \frac{2}{\Delta}\ln[\frac{|\Delta - 1|}{\Delta + 1}],
{}~~ for~~Q^{2}>M_{W}^{2}\cr}
\EQN ExplicitTheta$$
for space-like $Q^{2}$,
where $\Delta= \sqrt{|\frac{M^{2}_{W}}{Q^{2}}-1|}$,
and
$$
{\hat{\Theta}_{\gamma} =
\frac{2}{\Delta}\ln[\frac{|\Delta - 1|}{\Delta + 1}] ~~~~(5.22)
$$
for time-like $Q^{2}$, where
$\Delta= \sqrt{\frac{M^{2}_{W}}{|Q^{2}|}+1}$.

The double integral
$\Theta_{Z}$ can in principle be expressed in a closed form in terms
of Spence functions (see for example
\reference{Veltman}
G.~t'Hooft and M.~Veltman,
\journal Nucl. Phys.; B 153,365 (1979)
\endreference
), but this is of limited usefuleness for practical calculations.
Instead, we evaluated this integral numerically.
In {\bf Table1} and {\bf Table2} we record the final answer for
the sum
${\Delta\kappa}_{\gamma}^{P}= \Theta_{\gamma} + \Theta_{Z}$
The momentum variable $Q$ is defined as
$Q\equiv sign(Q^2)\sqrt{|Q^2|}$.
We used the same values for the constants
appearing in our calculations as in \cite{Lahanas},
namely $\alpha_{\gamma}=\frac{1}{128}$, $M_{W}=80.6 GeV$,
$M_{Z}=91.1 GeV$ and $s=0.23$ .
The overall numerical contribution
 of  $\Delta\kappa^{P}$ is rather
 small. Compared to the
typical contributions coming from gauge boson graphs in the
Feynman-t'Hooft gauge,
evaluated in \cite{Lahanas}, it is at least
one order of magnitude smaller. However, it is comparable to the
respective contributions from fermion graphs.
 The experimental
relevance of the additional contributions to $\Delta\kappa^{P}$,
which originate from pinching,
 will clearly depend
on the accuracy achieved in the next generations of experiments.
If we assume for a moment
 that the experimental precision could be such as to distinguish between
the values of \cite{Lahanas} and ours, the difference could easily be
interpreted (or mistaken) for the contribution of a 4th fermion family,
for example,
since as it turned out, pinch contributions,
although of a totally different origin, are
comparable to fermionic ones. Regardless of that however, from the
theoretical point of view, our results augment those of
\cite{Lahanas}, rendering them gauge fixing parameter independent
and infrared finite in a natural way.

\section {Conclusions}
In this paper we have
undertaken a study of the structure of the
trilinear gauge boson vertices in the context of the Standard Model.
Using the S-matrix pinch technique we
constructed
to one-loop order
gauge independent $\gamma WW$ and $ZWW$ vertices, with all three
incoming momenta off-shell.
In the limit $p_{1}^{2},p_{1}^{2} \rightarrow M^{2}_{W}$
the gauge-independent vertices give rise to expressions for the magnetic
dipole moment $\mu_{W}$ and electric quadrupole moment $Q_{W}$, which,
unlike previous treatments, are suitable for comparison with experimental
observation
\reference{DSM}
The general strategy of how to extract
 the experimental values of quantities
like $\Delta\kappa$ from the total
scattering cross-section has been
outlined in

G.~Degrassi, A.~Sirlin, and W.~Marciano,~~
\journal Phys. Rev.; D 39,287 (1989)
\endreference
. The main effect of the pinch contributions is to render the results
gauge independent and infrared finite, while their numerical contribution
to the final answer turned out
to be small.

The gauge independent off-shell $\gamma W^{+}W^{-}$ vertex was shown
to satisfy a simple QED-like Ward identity, which relates it to the
gauge-independent W self-energies introduced by Degrassi and Sirlin.
It would be interesting to determine whether or not the
gauge-independent $ZW^{+}W^{-}$ vertex satisfies a similar
Ward identity. Calculations in this direction are already in progress.

\section {Acknowledgements}
The authors are indebted to Professor A.~Sirlin for many useful
discussions, and to A.~Bunch and T.~Battacharya for technical support.
This work was supported
in part by the National Science Foundation under Grant No. PHY-9017585.
\section {References.}
\ListReferences
\section{Figure Captions}

1) The kinematics of the $VW^{+}W^{-}$ vertex. All momenta are incoming.

2) Graphs (a)-(c) are some of the contributions to the S-matrix $T$.
Graphs (e) and (f) are pinch parts, which, when added to the usual
self-energy graphs (d), give rise to a gauge-independent effective
self-energy.

3) The general structure of the part $T(q,p_{1},p^{2})$
of the S-matrix, that only depends on the momentum transfers.

4) The usual triangle graphs contributing to the
 $\gamma W^{+}W^{-}$ vertex,
referred to as $V_{\mu\alpha\beta}^{i}$ in the text.
All graphs are understood
to be sandwiched between external fermions, as in Fig.3.

5) The rest of the usual  $V_{\mu\alpha\beta}^{i}$
graphs contributing to the $\gamma W^{+}W^{-}$ vertex.

6) Graphs (a) and (c) contribute to the
gauge-independent vertex through their pinch parts,
graphs (b) and (d) respectively. Notice the absence of
box graphs containing $\gamma\gamma$, $ZZ$, or $Z\gamma$
legs.

7) The pinch contribution of the box containing a $\gamma$

8) The pinch contribution of the box containing a $Z$.
Notice that the direct and the crossed graphs are not
equivalent, since the internal fermion is an electron
or a neutrino, respectively

9) The Feynman graphs contributing to the gauge-independent
$WW$ self-energy. The last graph denotes the pinch contributions
(in the $\xi=1$ gauge).

\section {Table Captions}

1) The values of $\Delta\kappa^{P}$ for space-like $Q^{2}$.

2) The values of $\Delta\kappa^{P}$ for time-like $Q^{2}$.

\bye